\documentclass[aps,prb,showpacs,twocolumn,superscriptaddress]{revtex4}
\usepackage[latin1]{inputenc}
\usepackage[T1]{fontenc}
\usepackage[french,dutch,english]{babel}
\usepackage{amsmath}
\usepackage{amssymb,amsfonts,textcomp}
\usepackage{color}
\usepackage{array}
\usepackage{hhline}
\usepackage{graphicx}
\usepackage{hyperref}
\usepackage{float}

\newcommand{\angstrom}{\textup{\AA}}
\setcitestyle{super}

\makeatletter
\newcommand\ps@Standard{
  \renewcommand\@oddhead{}
  \renewcommand\@evenhead{}
  \renewcommand\@oddfoot{}
  \renewcommand\@evenfoot{\@oddfoot}
  \renewcommand\thepage{\arabic{page}}
}
\makeatother

\begin{document}
\title{Quasiparticle Interference on the Surface of Topological Crystalline Insulator Pb$_{1-x}$Sn$_x$Se}

\author{A.~Gyenis}
\author{I.~K.~Drozdov}
\author{S.~Nadj-Perge}
\author{O.~B.~Jeong}
\author{J.~Seo}
\affiliation{Joseph Henry Laboratories and Department of Physics, Princeton University, Princeton, New Jersey 08544}

\author{I.~Pletikosi\'{c}}
\affiliation{Joseph Henry Laboratories and Department of Physics, Princeton University, Princeton, New Jersey 08544}
\affiliation{Condensed Matter Physics and Materials Science Department, Brookhaven National Lab, Upton, New York 11973}

\author{T.~Valla}
\author{G.~D.~Gu}
\affiliation{Condensed Matter Physics and Materials Science Department, Brookhaven National Lab, Upton, New York 11973}

\author{A.~Yazdani}
\thanks{Corresponding author}
\affiliation{Joseph Henry Laboratories and Department of Physics, Princeton University, Princeton, New Jersey 08544}

\date{\today}
\begin{abstract}
Topological crystalline insulators represent a novel topological phase of matter in which the surface states are protected by discrete point group-symmetries of the underlying lattice. 
Rock-salt lead-tin-selenide alloy is one possible realization of this phase which undergoes a topological phase transition upon changing the lead content. We used scanning tunneling 
microscopy (STM) and angle resolved photoemission spectroscopy (ARPES) to probe the surface states on (001) Pb$_{1-x}$Sn$_{x}$Se in the topologically non-trivial (x=0.23) and 
topologically trivial (x=0) phases. We observed quasiparticle interference with STM on the surface of the topological crystalline insulator and demonstrated that the 
measured interference can be understood from ARPES studies and a simple band structure model. Furthermore, our findings support the fact that Pb$_{0.77}$Sn$_{0.23}$Se and PbSe have different topological nature.
\end{abstract}

\maketitle
 
\subsection{I.~\textsc{Introduction}}
In condensed matter physics, the study of topological phenomena has been in the focus of the research for the past few years. After the theoretical prediction and experimental observation 
of $\mathrm{Z}_{2}$ topological insulators\cite{ti1,ti2,ti3,ti4,ti5,ti6,ti7,ti8,ti9,ti10,ti11,ti12,ti13}, more recently, a new phase called topological crystalline insulator (TCI) has been proposed\cite{LiangFu}. Unlike the widely studied topological insulators, TCIs have an \textit{even} number of band inversions, which makes them trivial 
under $\mathrm{Z}_{2}$ classification. Nevertheless, due to the presence of crystal symmetry, these materials still have topologically protected surface states.

The first theoretically proposed TCI was a IV-VI semiconductor, SnTe\cite{Hsieh}. Its topologically non-trivial nature arises from the mirror symmetry present in its rock-salt crystal structure and the even number of band inversions.  
By substituting the Sn content with Pb, the strength of the spin-orbit coupling can be tuned which leads to a non-inverted band structure\cite{PbSn1,PbSn2}. Thus, Pb$_{1-x}$Sn$_{x}$Te and 
similarly, Pb$_{1-x}$Sn$_{x}$Se have a topological phase transition as a function of the doping level. The even number of Dirac cones of the TCI phase has been already observed in a number of angle resolved photoemission spectroscopy (ARPES) studies conducted on Pb$_{1-x}$Sn$_{x}$Se, Pb$_{1-x}$Sn$_{x}$Te and SnTe compounds\cite{PbSnSe,PbSnTe,SnTe}.
 
 We carried out spectroscopic measurements with a scanning tunneling microscope to study the quasiparticle interference (QPI) on the (001) surface 
of Pb$_{1-x}$Sn$_x$Se. We show that the QPI patterns are the direct consequence of the scattering between the electronic band pockets observed by ARPES. 
We also use a theoretical model to demonstrate that the trends in the energy-momentum dispersion of the QPI peaks can be understood within a simple framework of joint density of states.
\begin{figure}[H]
\centering
\includegraphics[width=0.4\textwidth]{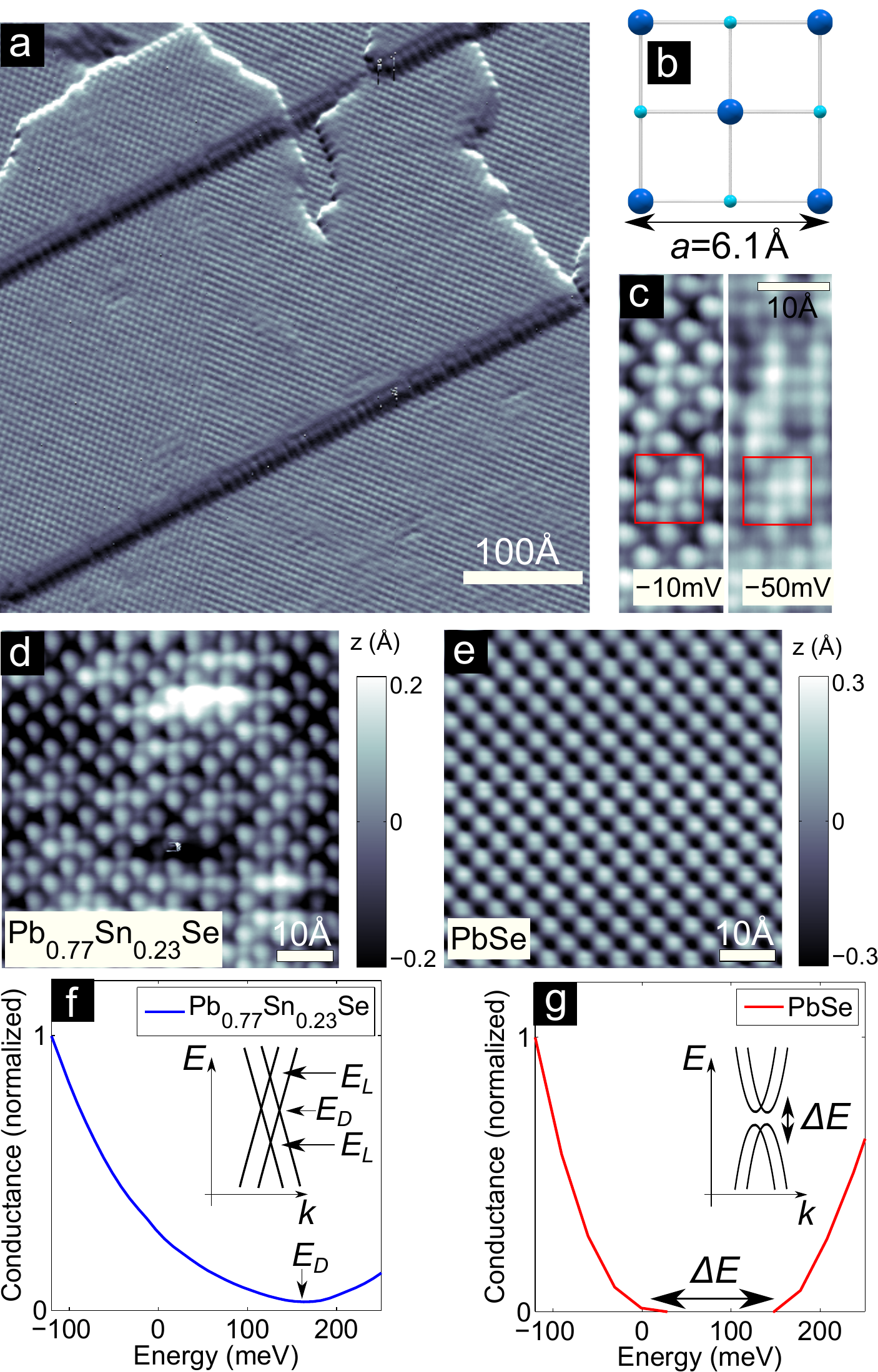}
\caption{a) STM topographic image ($V_\textrm{bias}=$400~mV and $I=25$~pA) of a 450~$\angstrom$-by-450~$\angstrom$ cleaved surface of  Pb$_{0.77}$Sn$_{0.23}$Se. b) Schematic illustration of the  (001) termination of the rocksalt crystal structure. (c) High spatial resolution topographic image of the same area at $V_\textrm{bias}=-10$~meV ($I=150$~pA) and at $V_\textrm{bias}=-50$~meV ($I=200$~pA). Topographies (d)-(e) and spatially averaged $dI/dV$ spectra (f)-(g) on Pb$_{0.77}$Sn$_{0.23}$Se and undoped PbSe. Insets show the schematic surface band dispersion as a function of the momentum along ${\mathrm{\bar{\Gamma}\bar{X}}}$ direction.
}
\label{fig1}
\end{figure}
\subsection{II.~\textsc{Results and Discussion}}
We studied single crystals of Pb$_{0.77}$Sn$_{0.23}$Se and PbSe, which were cleaved in ultra-high vacuum at room temperature. The STM measurements were performed in the temperature range of $T=30-50$~K using a home-built cryogenic STM. The ARPES experiments were carried out at the U13UB beamline of the National Synchrotron Light Source with the 18~eV photons. The electron analyzer was a Scienta SES-2002 with the combined energy resolution around 8 meV and the angular resolution of $\sim 0.15^\circ$. 

The STM topographic images of both Pb$_{0.77}$Sn$_{0.23}$Se and PbSe (Fig.~\ref{fig1}a-e) reveal that the cleaving process leads to atomically flat regions separated by single atomic steps of half unit cell height ($a=6.1~\angstrom$). This 
confirms that the cleaving process indeed exposed the (001) surface of the crystal. The two fcc sublattices (Pb/Sn and Se) can be separately imaged by changing the sample bias, which was observed in case of both samples. For example, in Pb$_{0.77}$Sn$_{0.23}$Se at the bias of -10~mV  only one 
sublattice is revealed, while at -50~mV both sublattices can be seen distinctly. We note that the possibility of observing both lattices and the bias voltage where the contrast reversal happens depends on the state of the STM tip. The Sn dopants can be identified as light dots (Fig.~\ref{fig1}d), which are obviously missing on the undoped sample (Fig.~\ref{fig1}e). Based on the location of the dopants, the sublattice observed at -10~mV is Se.

Spectroscopic ($dI/dV$) measurements show that while in the case of the topologically trivial (non-TCI) PbSe sample, there is a well-defined gap of 120~meV in the spectra, the density of states of the non-trivial TCI sample has a pronounced minimum but no gap (Fig.~\ref{fig1}f-g). Since PbSe is a trivial insulator, the surface states can be gapped, which is consistent with the observed spectrum measured by STM and further confirmed by our ARPES measurements (not shown). In the case of the TCI sample, however, the topological protection guaranties the existence of the metallic surface states at all energy values, and the Dirac dispersion leads to a minimum in the density of states at $E_D$=160~meV Dirac energy (Fig.~\ref{fig1}f-g insets).  The position of the Fermi level ($E=0$) indicates the p-type character of the samples, which is consistent with our ARPES results. It has been predicted that the topology of the Fermi surface changes as a function of energy (Lifshitz transition)\cite{Lifshitz, Hsieh}. In our measurements, however, no singularities arising from the Lifshitz transition points ($E_L$) have been observed in point spectroscopy either due to the overlap of the surface band with the bulk states or disorder smearing of the van-Hove singularities. The presence of intrinsic disorder in the studied samples, however, allows us to visualize the scattering processes on the surface relevant for potential future device applications of these materials.

\begin{figure}[H]
\centering
\includegraphics[width=0.45\textwidth]{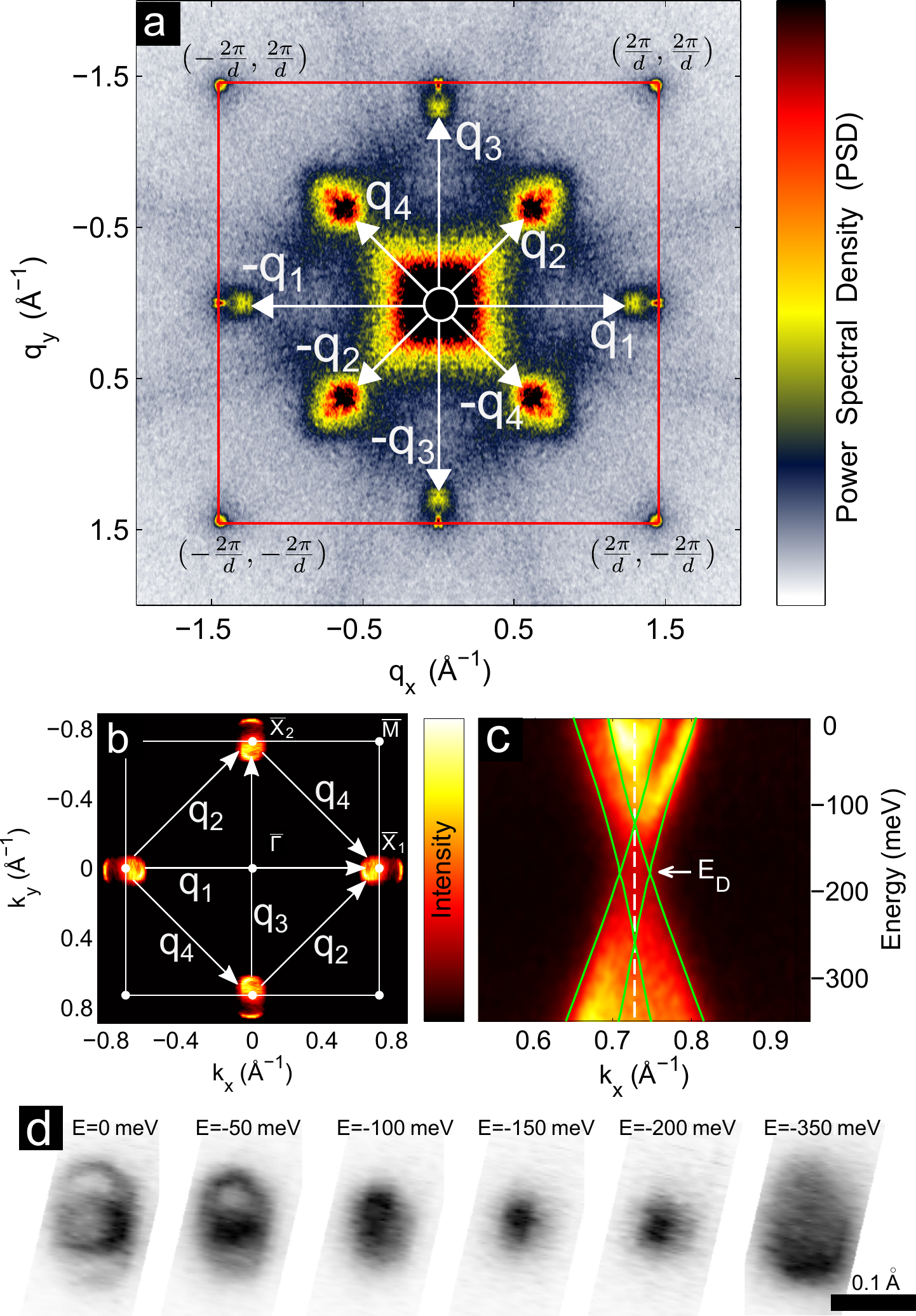}
\caption{(a) Fourier transform of a conductance map taken at $V_{\mathrm{bias}}=-100$~meV and $I=40$~pA over an area of 820$\angstrom\times$820$\angstrom$. Red lines mark the boundary of the FSBZ. (b) ARPES intensity map at $E=-100$~meV on Pb$_{0.77}$Sn$_{0.23}$Se reveals overall four surface pockets in the inner side and four other pockets on the outer side of $\mathrm{\bar{X}}$ points. White square indicates the location of the FBZ. (c) Energy-momentum dispersion relation measured by ARPES on Cs doped Pb$_{0.85}$Sn$_{0.15}$Se in the $\mathrm{\bar{\Gamma}\bar{X}}$ direction. Green lines show the highest intensity obtained from the theoretical model. (d) ARPES intensity maps of Cs doped Pb$_{0.85}$Sn$_{0.15}$Se around the $\mathrm{\bar{X}_{2}}$ point.}
\label{fig2}
\end{figure}

In order to obtain information about the scattering within the surface states, Fourier transform scanning tunneling spectroscopy (FT-STS) technique was used.  By measuring real-space variations in the differential conductance maps induced by disorder,  one can obtain energy and momentum-resolved information about the scattering processes happening at the surface of the material. At a certain energy, a \textbf{q}-wavevector modulation in the local density of states corresponds to the interference of the quasiparticles at momentum $\textbf{k}_{1}$ and $\textbf{k}_{2}$, satisfying $\textbf{q}=\textbf{k}_{1}$-$\textbf{k}_{2}$. Since in an FT-STS experiment we observe the differences  between $\textbf{k}_{1}$ and $\textbf{k}_{2}$ wavevectors, the signal coming only from the first Brillouin zone (FBZ) can be described in  $\textbf{q}$ scattering space with a zone twice the dimensions of the FBZ, which we further refer to as first scattering Brillouin zone (FSBZ).
In case of this study, the FSBZ (marked as red box on Fig.~\ref{fig2}a) is a square with size of $\frac{4\pi}{d}\times\frac{4\pi}{d}$, where $d=a/\sqrt{2}$ and oriented in the same way as the FBZ: $\mathbf{q}_{\mathrm{x}}$ ($\mathbf{q}_{\mathrm{y}}$) is parallel to the $\mathrm{\bar{\Gamma}\bar{X}_{1}}$ ($\mathrm{\bar{\Gamma}\bar{X}_{2}}$) direction.
Fig.~\ref{fig2}a~shows the Fourier transform of a conductance map obtained on the surface of Pb$_{0.77}$Sn$_{0.23}$Se, which reveals many pronounced wavevectors. The outer eight sharp features, which lie on the boundary of the FSBZ, correspond to the atomic structure. The existence of these peaks (Bragg peaks) is the result of the inevitable fact that the conductance measurement is performed on the atomic lattice, therefore, the map will include modulation arising from the atomic corrugation. The broader and more pronounced wavevectors (marked as $\pm \mathbf{q_1}$ to $\pm \mathbf{q_4}$) reside inside the FSBZ ($\pm \mathbf{q_1}$ and $\pm \mathbf{q_3}$ touch the Bragg-peaks), and correspond to interband scattering between the different pockets of the constant-energy surface. Finally, the central peak comes from long-wavelength modulations due to disorder and intraband scattering contributions.

Since the system possesses mirror and rotational symmetry, there are only two inequivalent wavevectors ($\mathbf{q_1}$ and $\mathbf{q_2}$) present in the scattering pattern. To identify the origin of these scattering wavevectors, we performed ARPES measurements on the samples. Similarly to previous studies\cite{PbSnSe,PbSnTe,SnTe}, the Fermi surface mapping  (Fig.~\ref{fig2}b) reveals two pockets on the two sides of the $\mathrm{\bar{X}}$ points. Based on this band structure information one can conclude that $\mathbf{q_1}$ and $\mathbf{q_2}$  scattering wavevectors observed in the STM experiment correspond to the interband scattering between the pockets in  $\mathrm{\bar{\Gamma}}\mathrm{\bar{X}_{1}}$ and $\mathrm{\bar{X}_{1}}\mathrm{\bar{X}_{2}}$ ($\mathrm{\bar{\Gamma}}\mathrm{\bar{M}}$) directions, respectively. The fact that the $\mathbf{q_1}$ peak and the Bragg peak are well resolved confirms that the Dirac nodes are slightly shifted away from the $\mathrm{\bar{X}}$ points. Furthermore, it is important to note that in our QPI measurements we did not observe large $\mathbf{q}$ vectors corresponding to scattering events between states located in the first and second Brillouin zone. This can be most clearly seen by looking at the $\mathrm{\bar{\Gamma}}\mathrm{\bar{X}_{1}}$ direction, where the $\mathbf{q_1}$ scattering wavevector is located entirely inside the FSBZ and no scattering intensity beyond the Bragg peak is detected.

A more quantitative understanding of the conductance maps can be achieved if we recall that the QPI pattern is closely related to the joint density of states (JDOS) of the surface electrons\cite{ti9,JDOS1,JDOS2,JDOS3}. JDOS at momentum difference \textbf{q}  and a certain energy $E$ is defined as an autoconvolution of initial and final densities of states: JDOS($\mathbf{q}$,$E$)=$\int{d^2\mathbf{k}\rho(\mathbf{k},E)\rho(\mathbf{k+q},E)}$. 

We compare the measured QPI pattern with a simple JDOS simulation, in which the momentum-space local density of states is obtained from an effective Hamiltonian derived from symmetry arguments in Ref.~\onlinecite{kp_model}. In general, the structure of the TCI surface states near the Dirac energy can be approximated by a four-band $\textbf{k}\cdot\textbf{p}$ model:
\begin{eqnarray}
\begin{split}
H=&m\Sigma_{30}+m'\Sigma_{10}+\left(v_{1x}\Sigma_{01}+v_{2x}\Sigma_{11}+v_{3x}\Sigma_{31}\right)k_{x} \\
&+\left(v_{1y}\Sigma_{03}+v_{2y}\Sigma_{13}+v_{3y}\Sigma_{33}\right)k_{y},
\end{split}
\end{eqnarray} 
where $\sigma_{\alpha}$ are the Pauli matrices and the Dirac matrices are defined as $\Sigma_{\alpha\beta}=\sigma_{\alpha}\otimes\sigma_{\beta}$. 
To find the parameters in the Hamiltonian, we used the band structure information obtained from ARPES. By doping the surface with Cs we were able to shift the surface states of the originally p-type sample by $\sim350$~meV and turn them into the n-type, with the Dirac points at $\sim190$~meV below the Fermi level. Cs was deposited from a
commercial (SAES) getter source while keeping the sample at $T\sim15$~K. 
Fig.~\ref{fig2}c shows the bandstructure of Cs doped Pb$_{0.85}$Sn$_{0.15}$Se sample measured along the $\mathrm{\bar{\Gamma}}\mathrm{\bar{X}}$ direction, and green lines show the highest intensity values obtained from the theoretical model with the following parameters: $m=-0.06$~eV, $m'=-0.03$~eV, $v_{1x}=-3.8~\mathrm{eV}\angstrom$, $v_{2x}=-1.5~\mathrm{eV}\angstrom$, $v_{3x}=0.003~\mathrm{eV}\angstrom$, $v_{1y}=0.003~\mathrm{eV}\angstrom$, $v_{2y}=0.003~\mathrm{eV}\angstrom$, $v_{3y}=-3.5~\mathrm{eV}\angstrom$. Lifetime effects were included by introducing $\delta=3$~meV broadening. Note that we did not observe any significant difference in the band structure between x=0.15 and x=0.23  Pb$_{1-x}$Sn$_{x}$Se samples (other than a small difference in the position of the Fermi level), which justifies that we can use the same parameters for both samples. For our sample the extracted mass terms and the Dirac velocity in \textit{y} direction are similar to those reported in the previous study\cite{wang}, but different parameters were used in the \textit{x} direction.
\begin{figure}[H]
\centering
\includegraphics[width=0.42\textwidth]{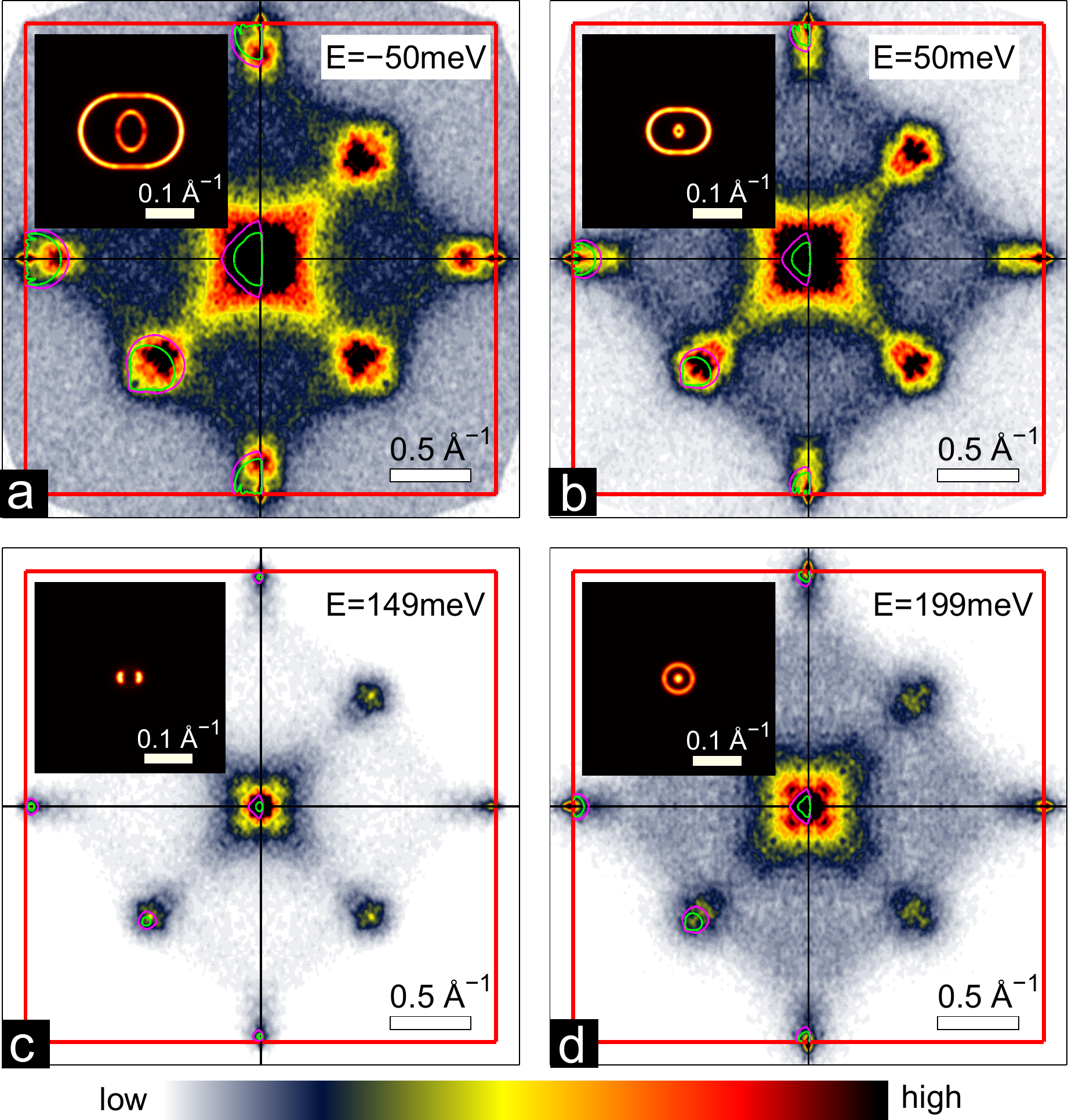}
\caption{Fourier transform of the QPI patterns on the surface of Pb$_{0.77}$Sn$_{0.23}$Se at different energies superimposed with the calculated JDOS (third and fourth quarter of \textit{q}-space). In the figures, two different isocontours are shown corresponding to two different intensity values: the intensity of the green contours is an order of magnitude higher than the intensity of purple contours. Insets display the corresponding calculated Fermi surfaces around the $\mathrm{\bar{X}}$ point.}
\label{fig3}
\end{figure}
Since we did not observe any scattering vector between the first and second Brillouin zone, we restricted the JDOS calculation to the FBZ. The resulting JDOS are overlayed with the Fourier transform of the differential conductance maps at different energies (Fig.~\ref{fig3}).
The same measurements were carried out on the trivial PbSe samples. As one would expect based on point spectroscopy measurements (Fig.~\ref{fig1}g), no surface state QPI peaks were observed on the maps obtained within the gap.
\begin{figure}[htb]
\centering
\includegraphics[width=0.5\textwidth]{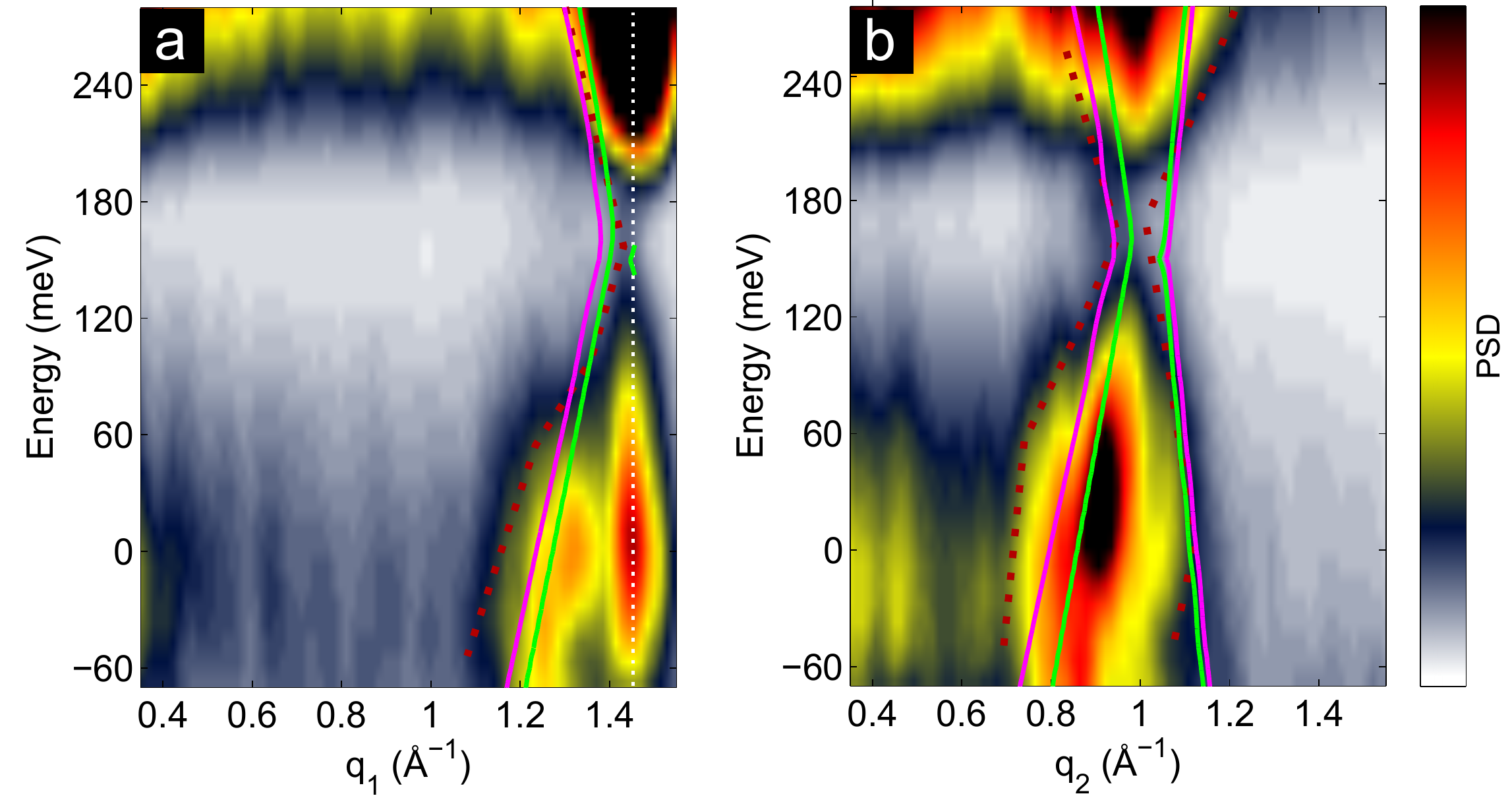}
\caption{Energy-momentum structure of the surface states of Pb$_{0.77}$Sn$_{0.23}$Se along (a) $\mathrm{\bar{\Gamma}}\mathrm{\bar{X}}$ and (b) $\mathrm{\bar{\Gamma}}\mathrm{\bar{M}}$ directions. White dashed line indicates the position of the Bragg peak, while red dotted lines as guides to the eye enclose the relevant high intensity regions of the STM data. The isocontours correspond to the same intensity values as on Fig.~\ref{fig3}.}
\label{fig4}
\end{figure}
Furthermore, Fig.~\ref{fig4} shows the QPI intensity as a function of energy and momentum along $\mathrm{\bar{\Gamma}}\mathrm{\bar{X}}$  and $\mathrm{\bar{\Gamma}}\mathrm{\bar{M}}$ scattering directions, respectively. In the $\mathrm{\bar{\Gamma}}\mathrm{\bar{X}}$ direction, one can observe two peaks: a non-dispersive peak (Bragg-peak) located at $\left|\mathbf{q}_{\mathrm{x}}\right|=2\pi/d$ and a slightly dispersive $\mathbf{q_{1}}$ peak. Above 60~meV $\mathbf{q_{1}}$ touches the  Bragg-peak and they are indistinguishable. In the $\mathrm{\bar{\Gamma}}\mathrm{\bar{M}}$ direction only the $\mathbf{q_{2}}$ is present. Both directions show the Dirac point around 160~meV, where the width of the peaks is the smallest. 

Although the theoretical model and our experimental findings show similar trends, we observe that there is a mismatch between the calculation and our measurement. The discrepancy between the lineshapes of the autoconvolved simulated DOS and STM data might be due to many factors. Such effects include energy-dependent quasiparticle lifetime broadening, inhomogeneous  broadening caused by disorder and the shape of the impurity potential or tip-induced band bending. Also, one should note that the intensity of the observed signal is rather weak to perform a more quantitative match to simulations.  Despite the fact that the JDOS model does not give an entirely satisfactory description of our experimental results, it is sufficient to capture the trend in the QPI dispersion cuts.  

We also note that we did not see evidence for wavevector suppression in our measurements. This is in contrast to the case of quasiparticle interference in topological insulators (previously studied Bi$_{1-x}$Sb$_{x}$\cite{ti9}, Bi$_2$(Se/Te)$_3$ \cite{hb}) or other materials with strong spin-orbit coupling (like Sb\cite{Sb_tmatrix1}), in which one has to invoke spin selection rules in order to properly account for the spin texture of the surface states and reproduce experimental data. On TCI materials, prohibited scattering vectors have been proposed\cite{kp_model}, however no signatures of this protection have been observed in our experiment. The underlying reason could be that there is only a discrete set of points at which the scattering is fully protected. The contribution of these points to the overall spectral weight is small compared to other allowed scattering wavevectors nearby. Furthermore, the size of the pockets  makes it impossible to unambiguously resolve the effect of protection in our STM experiment.
\subsection{III.~\textsc{Conclusions}}
In conclusion, we studied quasiparticle interference on the surface of Pb$_{1-x}$Sn$_{x}$Se compounds using STM. We demonstrated that the observed QPI is directly related to the scattering of surface states between the four surface pockets measured by ARPES on the same samples. Our results support that the x=0 and x=0.23 compounds belong to two different topological classes, and Pb$_{0.77}$Sn$_{0.23}$Se is a non-trivial TCI with topologically protected surface states.
\subsection{IV.~\textsc{Acknowledgements}}
We thank B.~Andrei Bernevig and Chen Fang for the discussions. The work at Princeton University was supported by NSF-DMR1104612, NSF-MRSEC programs through the Princeton Center for Complex Materials (DMR-0819860), DARPA-SPAWAR grant N6601-11-1-4110, and ARO MURI program, grant W911NF-12-1-0461. The work at Brookhaven National Lab is supported by U.~S.~Department of Energy, Office of Basic 
Energy Sciences, under contract No.~DE-AC02-98CH10886. S.~N.-P.~acknowledges support of the European Community under a Marie-Curie OEF fellowship.


\begin{thebibliography}{24}

\bibitem{ti1}C.~L.~Kane and E.~J.~Mele, Phys.~Rev.~Lett.~\textbf{95}, 146802 (2005).
\bibitem{ti2}J.~E.~Moore and L.~Balents, Phys.~Rev.~B \textbf{75}, 121306(R) (2007).
\bibitem{ti3}L.~Fu, C.~L.~Kane, and E.~J.~Mele, Phys.~Rev.~Lett.~\textbf{98}, 106803 (2007).
\bibitem{ti4}C.~L.~Kane and E.~J.~Mele, Phys.~Rev.~Lett.~\textbf{95}, 226801 (2005).
\bibitem{ti5}B.~A.~Bernevig, T.~L.~Hughes and S.-C.~Zhang, Science \textbf{314}, 1757 (2006).
\bibitem{ti6}M.~Konig, S.~Wiedmann, C.~Brune, A.~Roth, H.~Buhmann, L.~W.~Molenkamp, X.-L.~Qi and S.-C.~Zhang, Science \textbf{318}, 766 (2007).
\bibitem{ti7}D.~Hsieh, D.~Qian, L.~Wray, Y.~Xia, Y.~S.~Hor, R.~J.~Cava and M.~Z.~Hasan, Nature \textbf{452}, 970 (2008).
\bibitem{ti8}D.~Hsieh, Y.~Xia, L.~Wray, D.~Qian, A.~Pal, J.~H.~Dil, J.~Osterwalder, F.~Meier, G.~Bihlmayer, C.~L.~Kane, Y.~S.~Hor, R.~J.~Cava and M.~Z.~Hasan, Science \textbf{323}, 919 (2009).
\bibitem{ti9}P.~Roushan, J.~Seo, C.~V.~Parker, Y.~S.~Hor, D.~Hsieh, D.~Qian, A.~Richardella, M.~Z.~Hasan, R.~J.~Cava and A.~Yazdani, Nature \textbf{460}, 1106 (2009).
\bibitem{ti10}J.~Seo, P.~Roushan, H.~Beidenkopf, Y.~S.~Hor, R.~J.~Cava and A.~Yazdani, Nature \textbf{466}, 343 (2010).
\bibitem{ti11}J.~E.~Moore, Nature \textbf{464}, 194 (2010).
\bibitem{ti12}S.-Y.~Xu, Y.~Xia, L.~A.~Wray, S.~Jia, F.~Meier, J.~H.~Dil, J.~Osterwalder, B.~Slomski, A.~Bansil, H.~Lin, R.~J.~Cava and M.~Z.~Hasan, Science \textbf{332}, 560 (2011).
\bibitem{ti13}X.-L.~Qi and  S.-C.~Zhang, Rev.~Mod.~Phys.~\textbf{83}, 1057 (2011).
\bibitem{LiangFu}L.~Fu, Phys.~Rev.~Lett.~\textbf{106}, 106802 (2011).
\bibitem{Hsieh}T.~H.~Hsieh, H.~Lin, J.~Liu, W.~Duan, A.~Bansil and L.~Fu, Nature Communications \textbf{3}, 982 (2012).
\bibitem{PbSn1}A.~J.~Strauss,  Phys.~Rev.~\textbf{157}, 608 (1967).
\bibitem{PbSn2}J.~O.~Dimmock, I.~Melngailis and A.~J.~Strauss, Phys.~Rev.~Lett.~\textbf{26}, 1193 (1966).
\bibitem{PbSnSe}P.~Dziawa, B.~J.~Kowalski,	K.~Dybko, R.~Buczko, A.~Szczerbakow, M.~Szot, E.~Lusakowska, T.~Balasubramanian, B.~M.~Wojek, M.~H.~Berntsen, O.~Tjernberg and T.~Story, Nature Materials \textbf{11}, 1023 (2012). 
\bibitem{PbSnTe}S.-Y.~Xu, C.~Liu, N.~Alidoust, M.~Neupane, D.~Qian, I.~Belopolski, J.~D.~Denlinger, Y.~J.~Wang, H.~Lin, L.~A.~Wray, G.~Landolt, B.~Slomski, J.~H.~Dil, A.~Marcinkova, E.~Morosan, Q.~Gibson, R.~Sankar, F.~C.~Chou, R.~J.~Cava, A.~Bansil and M.~Z.~Hasan, Nature Communications \textbf{3}, 1192 (2012).
\bibitem{SnTe}Y.~Tanaka, Z.~Ren, T.~Sato, K.~Nakayama, S.~Souma, T.~Takahashi, K.~Segawa and Y.~Ando, Nature Physics \textbf{8}, 800 (2012).
\bibitem{Lifshitz}I.~M.~Lishitz, Zh.~Exp.~Teor.~Fiz., \textbf{38}, 1565 (1960) (Sov.~Phys.~JETP \textbf{11}, 1130 (1960)).
\bibitem{JDOS1}J.~E.~Hoffman, K.~McElroy, D.-H.~Lee, K.~M.~Lang, H.~Eisaki, S.~Uchida and J.~C.~Davis, Science \textbf{297}, 1148 (2002).
\bibitem{JDOS2}Q.-H.~Wang and D.-H.~Lee, Phys.~Rev.~B \textbf{67}, 20511 (2003).
\bibitem{JDOS3}R. S.~Markiewicz, Phys.~Rev.~B \textbf{69}, 214517 (2004).
\bibitem{kp_model}C.~Fang, M.~J.~Gilbert, S.-Y.~Xu, B.~A.~Bernevig and M.~Z.~Hasan, arXiv:1212.3285 [cond-mat.mes-hall] (2012).
\bibitem{wang}Y.~J.~Wang, W.~F.~Tsai, H.~Lin, S.-Y.~Xu, M.~Neupane, M.~Z.~Hasan and A.~Bansil , arXiv:1304.8119 [cond-mat.mtrl-sci] (2013).
\bibitem{hb}H.~Beidenkopf, P.~Roushan, J.~Seo, L.~Gorman, I.~K.~Drozdov, Y.~S.~Hor, R.~J.~Cava and A.~Yazdani, Nature Physics \textbf{7}, 939 (2011).
\bibitem{Sb_tmatrix1}A.~Strozecka, A.~Eiguren, M.~Bianchi, D.~Guan, C.~H.~Voetmann, S.~Bao, P.~Hofmann and J.~I.~Pascual, New J.~Phys.~\textbf{14}, 103026 (2012). 

\end{thebibliography}
\end{document}